\documentstyle[12pt]{article}
\title{Inappropriateness of the Rindler quantization}
\author{Hrvoje Nikoli\'c \\
Theoretical Physics Division, Rudjer Bo\v{s}kovi\'{c} Institute, \\
P.O.B. 180, HR-10002 Zagreb, Croatia \\
{\normalsize hrvoje@faust.irb.hr} \\
\makebox[1in]{} \\
}
\date{\today}
\begin{document}
\maketitle
\begin{abstract}
It is argued that the Rindler quantization is not a correct approach 
to study the effects of acceleration on quantum fields. 
First, the ``particle"-detector approach based on the Minkowski 
quantization is not equivalent to the approach based on 
the Rindler quantization.
Second, the event horizon, which plays the essential role
in the Rindler quantization, cannot play any physical role for a local
noninertial observer. 
\end{abstract}

\noindent
There is a wide belief that the properties of a quantum field 
seen by an uniformly accelerated observer are correctly 
described by the Rindler quantization, i.e., 
by the quantization 
based on the Rindler coordinates. On the other hand,   
it is known that the Rindler quantization suffers from certain 
problems. For example, 
the Rindler quantization is unitarily inequivalent to the Minkowski
quantization. However, this fact, being an artefact of the infinite volume
\cite{gerlach}, is only a technical problem which is not really serious.
Recently, it has been argued 
that the Rindler quantization suffers from a more serious problem
\cite{fedotov}, namely, 
that the boundary condition on the horizon, required by the Rindler
quantization, actually makes the Rindler quantization inconsistent
outside the left and right wedges. 

When quantization based od Rindler coordinates was discovered
\cite{full}, it was argued that it could be appropriate 
to the physical situation of an impenetrable wall located 
on the horizon.   
However, in \cite{unruh} and most other papers that apply the 
Rindler quantization it is assumed that the horizon itself 
plays a physical role because it serves as 
a physical boundary that affects the properties of quantum fields. 
In this letter we argue that such an assumtion is groundless. 

There is also a wide belief \cite{bd} that the Unruh effect 
\cite{unruh}, i.e., the thermal properties of the Minkowski 
vacuum seen by an uniformly accelerated observer, can be 
equivalently described 
with the Minkowski quantization, using a model of a 
``particle" detector. In this letter we show 
that the ``particle"-detector approach to the Unruh
effect based on the Minkowski quantization 
is not equivalent to the Rindler-quantization approach.     

Let us start with the discussion of a ``particle" detector. For
definiteness, we use the model of a monopole detector described in
\cite{bd}. Assuming that the detector and the field are in the ground
state $|0,E_0\rangle$ initially, the first order of perturbation    
theory gives the amplitude for the
transition to an excited state $|k,E\rangle$:
\begin{equation}\label{1}
A(k,\Delta E)=\bar{g} \int_{-\infty}^{\infty} d\tau \, e^{i\Delta E\,\tau}
\langle k|\phi(x(\tau))|0\rangle \; , 
\end{equation}
where $\bar{g}=ig\langle E|m(0)|E_0\rangle$, $g$ is a
real dimensionless coupling constant, 
$m(\tau)$ is the monopole moment operator, $x(\tau)$ is the trajectory
of the detector, $\Delta E=E-E_0$, and  
\begin{equation}\label{2}  
\langle k|\phi(x)|0\rangle =\frac{1}{\sqrt{(2\pi)^3 2\omega}} \,  
e^{i(\omega t -\bf{k}\cdot\bf{x})} \; .
\end{equation}
We compare the predictions that can be obtained from this model 
based on the Minkowski quantization 
with the predictions that result from the Rindler quantization
\cite{unruh2,muller}. 

The 
Rindler-quantization approach predicts that the absorption
of a Rindler particle by the accelerated atom will be seen 
by an inertial observer as an emission of a Minkowski particle.   
These are two descriptions of the same event seen by two 
different observers. According to this interpretation, the 
accelerated observer does not observe the emission of the Minkowski
particle.
On the other hand, in the approach based on (\ref{1}), both
the inertial and the accelerated observer can observe both
the jump to a higher atom level {\em and} the emission of the Minkowski   
particle, as two different events. Obviously, the two approaches 
are not equivalent.    

For a uniform 
acceleration, the two approaches agree in the prediction of a thermal
distribution for $\Delta E$. However, even this partial agreement of the two
approaches does not generalize  
when the uniform acceleration is replaced by
a more complicated motion \cite{padm3}. 

We now see that at least one of the two approaches
must be wrong. Below we argue that it is the Rindler quantization that is
wrong. 

Even if one does not regard the mentioned technical problems 
\cite{gerlach,fedotov} with the Rindler quantization  
as really serious problems, one cannot deny
that the event horizon plays the essential role for understanding
the physical consequences of the Rindler quantization.
Below we show that the event horizon does not correspond 
to any physical entity that could influence the properties of the fields
seen
by an accelerated observer, making the Rindler quantization physically
meaningless. 

Let $x'$ be the Fermi coordinates of an observer at $\mbox{\bf{x}}'=\bf{0}$
moving arbitrarily in flat spacetime.
If the observer does not rotate, the corresponding metric 
is given by $g'_{ij}=-\delta_{ij}$, $g'_{0i}=0$, and \cite{nels,nikolic1}
\begin{equation}\label{1.1}
g'_{00}(t',\mbox{\bf{x}}')=(1+\mbox{\bf{a}}'(t')\cdot\mbox{\bf{x}}')^2 \; ,
\end{equation}
where $\mbox{\bf{a}}'$ is the proper acceleration. From (\ref{1.1}) we see
that
the Fermi coordinates of an accelerated observer possess a coordinate
singularity at a certain $\bf{x}'$. However, in general, this coordinate
singularity does not correspond to any physical boundary. Only
$\bf{a}'(\infty)$, defining the event horizon, defines a physical 
boundary. However, in real life, acceleration never lasts infinitely long.
And even if it does, the event horizon
does not have any physical influence on a measuring
procedure that lasts a finite time. On the other hand, a realistic 
measuring procedure always lasts a finite time. Therefore, the 
horizon cannot play any physical role. 
 
Actually, the correct interpretation
of the Fermi coordinates, and therefore also of the Rindler coordinates
as their special case, is purely local \cite{nikolic1,nikolic2}, so
they are not appropriate for quantization which requires a global approach
to describe the EPR-like correlations. 

We do not see a reason to doubt that the Minkowski quantization is the 
correct approach to quantization in a flat background. In particular, 
it is not in contradiction with the principle of general covariance. 
The only problem, not yet satisfactorily solved, is how to generalize 
it to a curved background.  

\section*{Acknowledgment}
This work was supported by the Ministry of Science and Technology of the
Republic of Croatia under Contract No. 00980102.


\begin{thebibliography}{99}
\bibitem{gerlach}
U. H. Gerlach, {\it Phys. Rev.} {\bf D40}, 1037 (1989).
\bibitem{fedotov}
A. M. Fedotov, V. D. Mur, N. B. Narozhny, V. A. Belinskii and
B. M. Karnakov, {\it Phys. Lett.} {\bf A254}, 126 (1999).  
\bibitem{full}
S. A. Fulling, {\it Phys. Rev.} {\bf D7}, 2850 (1973).
\bibitem{unruh}
W. G. Unruh, {\it Phys. Rev.} {\bf D14}, 870 (1976).     
\bibitem{bd}  
N. D. Birrell and P. C. W. Davies, {\it Quantum fields in
curved space} (Cambridge Press, NY, 1982).
\bibitem{unruh2}
W. G. Unruh and R. M. Wald, {\it Phys. Rev.} {\bf D29}, 1047 (1984).
\bibitem{muller}
J. Audretsch and R. M\"{u}ller, {\it Phys. Rev.} {\bf D49}, 6566 (1994).
\bibitem{padm3}
L. Sriramkumar and T. Padmanabhan, preprint gr-qc/9903054.
\bibitem{nels}  
R. A. Nelson, {\it J. Math. Phys.} {\bf 28}, 2379 (1987). 
\bibitem{nikolic1}
H. Nikoli\'c, {\it Phys. Rev.} {\bf A61}, 032109 (2000).
\bibitem{nikolic2}
H. Nikoli\'c, preprint gr-qc/9909035.  

\end{thebibliography}
\end{document}